\documentclass[floats,floatfix,amssymb,prl,twocolumn,nofootinbib,superscriptaddress]{revtex4-2}
\usepackage[utf8]{inputenc}
\usepackage{amsmath,amssymb}
\usepackage{amsfonts}
\usepackage{bm}
\usepackage{courier}
\usepackage{graphics,orcidlink}
\usepackage{varioref}
\usepackage{mathtools}
\usepackage[normalem]{ulem} 
\usepackage{acronym}
\usepackage[caption=false]{subfig}
\usepackage{lipsum}
\usepackage{multirow}
\usepackage{graphicx}
\usepackage{tensor}
\usepackage{verbatim}
\usepackage{wrapfig}
\usepackage{nccmath}
\usepackage{soul} 
\allowdisplaybreaks

\labelformat{section}{Section #1} 
\labelformat{subsubsection}{Section #1}
\labelformat{subsubsubsection}{Section #1}
\labelformat{equation}{Eq.~(#1)} 
\labelformat{figure}{Fig.~#1} 
\labelformat{subfigure}{Fig.~\thefigure#1} 
\labelformat{table}{Tab.~#1} 
\labelformat{appendix}{Appendix #1}

\begin{document}

\title{Fermionic Love of Black Holes in General Relativity}

\author{Sumanta Chakraborty\orcidlink{0000-0003-3343-3227}}
\email{tpsc@iacs.res.in}
\affiliation{School of Physical Sciences, Indian Association for the Cultivation of Science, Kolkata-700032, India}

\author{Pierre Heidmann\orcidlink{0000-0003-3423-4115}}
\email{heidmann.5@osu.edu}
\affiliation{Department of Physics and Center for Cosmology and AstroParticle Physics (CCAPP), The Ohio State University, Columbus, OH 43210, USA}

\author{Paolo Pani\orcidlink{0000-0003-4443-1761}}
\email{paolo.pani@uniroma1.it}
\affiliation{Dipartimento di Fisica, ``Sapienza'' Universit\`a di Roma \& Sezione INFN Roma1, P.A. Moro 5, 00185, Roma, Italy}

\begin{abstract}
Black holes in General Relativity exhibit a remarkable feature: their response to static scalar, electromagnetic, and gravitational perturbations --- as quantified by the so-called tidal Love numbers --- vanishes identically.
We present the first exception to this rule: the Love numbers of a black hole perturbed by a fermionic field are nonzero. We derive a closed-form expression of these fermionic Love numbers for generic spin in the background of a Kerr black hole with arbitrary angular momentum. In contrast, we show that the fermionic dissipation numbers vanish for static perturbations, reflecting the absence of superradiance for fermions.
These results highlight a fundamental distinction between bosonic and fermionic perturbations, which can be interpreted as a breaking of the hidden symmetries that underlie the vanishing of Love numbers in the bosonic sector.
\end{abstract}
\maketitle
{\bf Introduction.}
Since the pioneering work of A.E.H.~Love, who analyzed Earth's deformations under the tidal forces of the Moon and the Sun~\cite{Love1909}, the response of self-gravitating bodies to external tidal fields has become a key probe of their internal structure~\cite{PoissonWill}.
This response is encapsulated in the tidal Love numbers, which quantify the induced multipole moments of the object generated by an external perturbing field.

In General Relativity, black holes exhibit a remarkable feature: their tidal Love numbers --- associated with the conservative response to static tidal fields --- vanish identically~\cite{Damour_tidal, Binnington:2009bb, Damour:2009vw, Gurlebeck:2015xpa,*Poisson:2014gka,*Landry:2015zfa, Pani:2015hfa, LeTiec:2020bos, Chia:2020yla,*LeTiec:2020spy, Bhatt:2023zsy}.
This result stands in sharp contrast with the behavior of other compact objects, which generically have nonzero tidal Love numbers~\cite{Cardoso:2017cfl}. This notably includes neutron stars --~for which tidal Love numbers measured from gravitational-wave mergers~\cite{Abbott:2018exr} play a crucial role in constraining the equation of state~\cite{Chatziioannou:2020pqz}~-- but also other exotic compact objects and black hole mimickers within and beyond General Relativity~\cite{Wade:2013hoa,Pani:2015tga,*Sennett:2017etc,*Mendes:2016vdr,*Uchikata:2016qku, *Raposo:2018rjn, *Cardoso:2019rvt,*Berti:2024moe,*Silvestrini:2025lbe,Cardoso:2017cfl,Chakraborty:2023zed}. Furthermore, the tidal Love numbers are nonzero also for black holes surrounded by matter distributions~\cite{Baumann:2018vus, *DeLuca:2021ite, *DeLuca:2022xlz, *Brito:2023pyl, *Capuano:2024qhv, *Cardoso:2019upw, *Cardoso:2021wlq,*DeLuca:2024uju,*Cannizzaro:2024fpz, *DeLuca:2025bph, Katagiri:2023yzm}, for black holes in modified gravity theories and asymptotically nonflat spacetimes~\cite{Cardoso:2017cfl, Cardoso:2018ptl, *DeLuca:2024nih, *DeLuca:2022tkm, *Barbosa:2025uau, *Barura:2024uog, *nair2024asymptotically-199, *Franzin:2024cah}, and finally for higher-dimensional black holes~\cite{Chakravarti:2018vlt, *Chakravarti:2019aup, *Pereniguez:2021xcj, *Dey:2020lhq, *Dey:2020pth, *Cardoso:2019vof, *Rodriguez:2023xjd, *Charalambous:2023jgq, *Charalambous:2024tdj, *Charalambous:2024gpf, *Singha:2025xah, Kol:2011vg, Hui:2020xxx, Ma:2024few}.

In the worldline effective field theory for compact objects, the tidal Love numbers appear as Wilson coefficients multiplying higher-derivative, finite-size operators and, at the leading quadrupole order, they scale as $\sim{\cal O}(1)R^5$, where $R$ is the typical size of the object and the prefactor depends on the object's internal structure.
Thus, the vanishing of black-hole tidal Love numbers in General Relativity poses a naturalness problem from the perspective of an effective field theory~\cite{Porto:2016zng}, suggesting an underlying symmetry that enforces such coefficients to vanish identically. 

Indeed, in recent years the vanishing of the black-hole tidal Love numbers  has been linked to hidden symmetries of perturbations of the Kerr solution in the zero-frequency limit~\cite{Hui:2020xxx, Charalambous:2021kcz, Hui:2021vcv, Berens:2022ebl, *BenAchour:2022uqo, Charalambous:2022rre, *Katagiri:2022vyz, *Ivanov:2022qqt, DeLuca:2023mio, Charalambous:2021mea, *Bonelli:2021uvf, *Ivanov:2022hlo, *Rai:2024lho, Bhatt:2023zsy, Sharma:2024hlz,*Lupsasca:2025pnt}.
These symmetry arguments have also been extended beyond the linear regime~\cite{DeLuca:2023mio, Riva:2023rcm, Iteanu:2024dvx}, and recently shown to persist to all orders in perturbation theory~\cite{Kehagias:2024rtz, Combaluzier-Szteinsznaider:2024sgb, Gounis:2024hcm} in the static case (while the dynamical tidal response of black holes has been studied in~\cite{Bhatt:2024yyz,Katagiri:2022vyz, Ivanov:2022qqt, DeLuca:2023mio,Charalambous:2022rre, Saketh:2023bul,Perry:2023wmm,Chakraborty:2023zed,Ivanov:2024sds,DeLuca:2024ufn,Bhatt:2024yyz,Katagiri:2024wbg,Katagiri:2024fpn,Bhatt:2024rpx,Caron-Huot:2025tlq,Chakraborty:2025wvs} with different motivations and approaches).

So far, these results have been limited to the \emph{bosonic} response of a black hole, while the \emph{fermionic} response is surprisingly unexplored. In this Letter, we close this missing piece and show that the static response of a Kerr black hole (including the nonrotating Schwarzschild limit) to a fermionic perturbation is nonzero, computing the corresponding fermionic Love numbers analytically. We use $G=c=1$ units.

{\bf Massless perturbations of a Kerr black hole.}
The unique stationary black hole solution to vacuum General Relativity is the Kerr metric~\cite{Kerr:1963ud,*Robinson}, characterized by only two parameters: the mass $M$ and the angular momentum $J=aM$, with $|a|\leq M$. The horizon radii are $r_\pm=M\pm\sqrt{M^2-a^2}$.
A remarkable property of this solution is that any perturbation can be decomposed into one or more spin-$s$ scalar wavefunctions $\Psi_s$ that admit a separable form,  
\begin{eqnarray}
    \Psi_s(t,r,\vartheta,\varphi) = R_s(r) S_s(\vartheta) \,e^{-i\omega t}e^{i m\varphi},
\end{eqnarray}
where $\{t,r,\vartheta,\varphi\}$ are Boyer-Lindquist coordinates, $\omega$ and $m$ denote the frequency and azimuthal number of the perturbation, and $s$ is its spin: integer for bosons and half-integer for fermions.  
The radial and angular functions, $R_s$ and $S_s$, are respectively governed by the \mbox{spin-$s$} Teukolsky equation and by a deformed spin-weighted spherical harmonic equation which, in the massless case, read~\cite{Teukolsky:1972my,*Teukolsky:1973ha,Teukolsky:1974yv}  
\begin{eqnarray}
 &&\mbox{\fontsize{10.4}{10.4}\selectfont\(\mathcal{L}_r R_s
+ \left( \frac{K^2 - 2is(r-M)K}{\Delta} + 4is\omega r - \lambda_s \right) \frac{R_s}{\Delta} = 0  ,\quad\)} \label{teuR}\\
 &&\mbox{\fontsize{10.4}{10.4}\selectfont\( \mathcal{L}_\theta S_s +\left[(s-a \omega \cos \vartheta)^2 - \frac{(m+s \cos \vartheta)^2}{\sin^2\vartheta} +C \right] S_s=0,\quad\)} \label{teuS}
\end{eqnarray}
where $\mathcal{L}_r \equiv \Delta^{-s-1} \frac{d}{dr} \left( \Delta^{s+1} \frac{d}{dr}\right)$, $\mathcal{L}_\theta \equiv \frac{1}{\sin\vartheta}\frac{d}{d\vartheta}\left(\sin\vartheta\frac{d}{d\vartheta}\right)$, $\Delta\equiv (r-r_+)(r-r_-)$, $\lambda_s$ the separation constant, and  $C\equiv s(1-s)+a\omega(2m-a \omega)+\lambda_s$. 

\vspace{0.1cm}
{\bf Static response of a Kerr black hole.} 
The ``susceptibility'' of a Kerr black hole to a massless spin-$s$ perturbation can be extracted from the large-distance expansion of the radial function in the static ($\omega=0$) limit~\cite{Chia:2020yla,LeTiec:2020bos,LeTiec:2020spy},  
\begin{align}\label{psi_4_intermediate}
R_s \underset{r\to \infty}{=} &\mathcal{E}_{s\ell m}\,r^{\ell-s} \Big[(1+\cdots)+{\cal F}_{s\ell m}\left(\frac{r_+}{r}\right)^{2\ell+1}(1+\cdots)\Big]\,,
\end{align}
where we have introduced the parameter $\ell$ such that 
\begin{equation}
    \lambda_s = \ell(\ell+1) - s(s+1),
\label{eq:RegAngularWave}
\end{equation}
and ${\cal F}_{s\ell m}$ encodes the response to the external field (the amplitude of which is set by $\mathcal{E}_{s\ell m}$), while the dots denote sub-leading terms.  
The real part of ${\cal F}_{s\ell m}$ defines the \emph{conservative response}, which is proportional to the tidal Love numbers, and its imaginary part encodes the \emph{dissipative response}, yielding the dissipation numbers~\cite{Goldberger:2005cd}.  

\vspace{0.1cm}
The angular equation for static perturbations reduces to the spin-weighted spherical harmonic equation~\cite{Goldberg:1966uu,*1977RSPSA.358...71B}, for which regularity at $\cos \vartheta=\pm 1$ requires $|m|\leq \ell$, $\ell\geq |s|$, and the quantization of $(\ell,m)$ in terms of $s$:
\begin{itemize}\vspace{-0.2cm}
    \item Bosonic perturbations: $(s,\ell,m)$ all integers;
    \vspace{-0.2cm}
    \item Fermionic perturbations: $(s,\ell,m)$ all half-integers.    
\end{itemize}\vspace{-0.2cm}
The corresponding regular waveforms are the spin-$s$ harmonics $S_s(\vartheta)e^{i m\varphi}={}_s Y_{\ell m}(\vartheta,\varphi)$~\cite{Goldberg:1966uu,1977RSPSA.358...71B}.  

\vspace{0.1cm}
{\bf A fermionic perturbation: Dirac fields.} While the response of the Kerr black hole to massless bosonic probes --- scalar ($s=0$), electromagnetic ($s=\pm 1$), and gravitational ($s=\pm 2$) fields --- has been extensively studied, one can also consider massless fermionic perturbations, such as Dirac fields ($s=\pm \tfrac{1}{2}$, e.g. neutrinos) \cite{Brill:1957fx,*ChandraBook,*Chandrasekhar:1976ap,Unruh:1973bda,Lee:1977gk} or Rarita-Schwinger fields ($s=\pm \tfrac{3}{2}$, e.g. gravitinos in supergravity)~\cite{Gueven:1980be,*TorresdelCastillo:1990aw}.

As an example, the dynamics of a massless spin-$\frac{1}{2}$ spinor is governed by the Dirac equation~\cite{Brill:1957fx,ChandraBook,Chandrasekhar:1976ap,Lee:1977gk},  
\begin{eqnarray}
\gamma^\mu\nabla_\mu\psi&=&0\,, \label{Dirac}
\end{eqnarray}
where $\left\{\gamma^\mu,\gamma^\nu\right\}=2 g^{\mu\nu}$,  $\nabla_\mu\psi=\partial_\mu\psi-\Gamma_\mu\psi$, $\Gamma_\mu$ is the spinor affine connection, and $g_{\mu\nu}$ the metric. In the Kerr background, the spinor admits a separable ansatz,
\begin{equation}
\medmath{
 \psi=\left(\frac{R_{-\frac{1}{2}}S_{-\frac{1}{2}}}{\sqrt{2}\rho^*},R_{+\frac{1}{2}}S_{+\frac{1}{2}},-R_{+\frac{1}{2}}S_{-\frac{1}{2}},-\frac{R_{-\frac{1}{2}}S_{+\frac{1}{2}}}{\sqrt{2}\rho}\right)^T e^{-i\omega t+im\varphi},} \nonumber
\end{equation}
with $\rho=r+ia\cos\vartheta$. 
The radial functions $R_{\pm\frac{1}{2}}(r)$ and angular functions $S_{\pm\frac{1}{2}}(\vartheta)$ are not independent and satisfy coupled first-order systems~(see Eqs.~(7-8) of \cite{Lee:1977gk}), 
which can be recast into decoupled spin-$\frac{1}{2}$ Teukolsky equations and deformed spherical harmonics, \ref{teuR} and \ref{teuS} with $s=\pm1/2$.

The radial waveforms are directly related to observable quantities: $|\sqrt{\Delta}\,R_{+\frac{1}{2}}|^2$ and $|R_{-\frac{1}{2}}|^2$ are directly related to the \emph{probability densities for spin-up and spin-down particles}. Moreover, although the Teukolsky equation~\ref{teuR} treats $R_{+\frac{1}{2}}$ and $R_{-\frac{1}{2}}$ as independent, they are related through the first-order system~\cite{Lee:1977gk}. In the static limit, this implies that the amplitudes $\mathcal{E}_{\pm 1/2\,\ell m}$ are related to each other, even though the corresponding spin-up and spin-down responses $\mathcal{F}_{\pm 1/2\,\ell m}$ may still differ.

{\bf Static tidal Love numbers of a Kerr black hole.}
For $\omega=0$,  Teukolsky's equation admits an analytic solution for arbitrary spin $s$:  
\begin{align}
    \medmath{R_s =}&\medmath{c_{1}\left(\frac{z}{1+z}\right)^{-iP}\,_{2}F_{1}(s-\ell,1+\ell+s;1+s-2iP;-z)} \nonumber \\
    &\medmath{+c_{2}\frac{(z(1+z))^{i P}}{z^s}\,_{2}F_{1}(-\ell+2iP,1+\ell+2iP;1-s+2iP;-z),}\nonumber
\end{align}
where $z\equiv (r-r_+)/(r_+-r_-)$ and $P\equiv am/(r_{+}-r_{-})$, which vanishes in the nonrotating limit.

Near the horizon, $z\to 0$, the solution behaves as  
\begin{align}
R_s\underset{z\to 0}{\sim}c_{1}z^{-iP}+c_{2}z^{-s} z^{iP}\,.
\end{align}
The first and second terms correspond to the zero-frequency limit of an outgoing and ingoing wave at the future horizon, respectively \cite{Teukolsky:1974yv}. Imposing ingoing boundary conditions requires $c_1=0$, which can also be directly inferred by imposing regularity of the static solution in the Hartle–Hawking tetrad, i.e. $\Delta^s R_s$ must be $C^2$ (twice continuously differentiable) at the horizon~\cite{LeTiec:2020bos}.  

At large $r$, the regular solution matches the generic form  in~\ref{psi_4_intermediate}, where we analytically continue $\ell\in \mathbb{C}$ for the expansion to be well-defined and to disentangle the response from the source~\cite{Kol:2011vg,Creci:2021rkz,Chia:2020yla,LeTiec:2020bos,LeTiec:2020spy}.  
\begin{widetext}
After some simplifications, the response function takes the form
\begin{align}
\mathcal{F}_{s\ell m}
&=\left(\frac{a m}{r_+} \right)^{2\ell+1}\times \frac{\Gamma(1+\ell-s)\Gamma(1+\ell+s)\,|\Gamma(1+\ell+2iP)|^2}{P^{2\ell+1}\,\Gamma(1+2\ell)\Gamma(2(1+\ell))}
\times \frac{\sin[\pi \ell-2i\pi P]\sin[\pi (\ell+s)]}{\pi \sin[2\pi \ell]}\,.
\end{align}
We now analytically continue back the response to integer and half-integer values of $\ell$, corresponding to bosonic and fermionic perturbations, respectively. After some algebra,\footnote{This simplification follows from expanding the sine factors and using standard identities involving Gamma functions with complex arguments, such as $\left| \Gamma(1+n+bi) \right|^2 
= \frac{\pi b}{\sinh(\pi b)} \prod_{k=1}^{n} \left(k^2 + b^2\right)$ and $\left| \Gamma\!\left(\tfrac{1}{2} + n + bi\right) \right|^2 
= \frac{\pi}{\cosh(\pi b)} \prod_{k=1}^{n} 
\left(\left(k-\tfrac{1}{2}\right)^2 + b^2\right)$.} this yields the main result of this Letter:
\begin{equation}
    \mathcal{F}_{s\ell m} = \begin{dcases}
        &i\,(-1)^{1-s}\,\frac{a m}{r_+}\,
\frac{(\ell-s)!\,(\ell+s)!}{(2\ell+1)!\,(2\ell)!}\,
\prod_{k=1}^{\ell} \left[ k^2 \left(1-\frac{r_-}{r_+} \right)^2 + \left(\frac{2 a m}{r_+}\right)^2 \right]\quad\hspace{0.3cm}\text{for bosons, $(s,\ell,m)$ integers}, \\
&\frac{(-1)^{\frac{1}{2} - s}}{2}\,
\frac{(\ell-s)!\,(\ell+s)!}{(2\ell+1)!\,(2\ell)!}\,
\prod_{k=1}^{\,\ell+\frac{1}{2}}
\left[ (k-\tfrac{1}{2})^2  \left(1-\frac{r_-}{r_+} \right)^2 + \left(\frac{2 a m}{r_+}\right)^2 \right] \quad \text{for fermions, $(s,\ell,m)$ half-integers},
    \end{dcases} \label{response}
\end{equation}
\end{widetext}

\emph{ Bosonic response.} For bosonic perturbations, i.e. $(s,\ell,m)$ integers, we retrieve that the response strictly vanishes for the Schwarzschild black hole as $\mathcal{F}_{s\ell m}^\text{Schw}=0$ when $a=0$~\cite{Damour_tidal, Binnington:2009bb, Damour:2009vw}.
For $a\neq0$, ${\cal F}_{s\ell m}$ is purely \emph{imaginary}. This corresponds to nonvanishing dissipation numbers for a spinning black hole, even for static perturbations, due to frame dragging~\cite{Chia:2020yla,LeTiec:2020bos,LeTiec:2020spy}. However, the conservative part of the response --- capturing the tidal Love numbers --- is strictly zero for any bosonic perturbations and any $a$.

{\emph{Fermionic response.}} The situation is markedly different for fermions, i.e. when $(s,\ell,m)$ are half-integers, including the Dirac field perturbations for $s=\pm \tfrac{1}{2}$. First, it should be noted that the fermionic responses are directly tied to physical observables, since the wavefunctions from which they are derived are associated with the probabilities of spin-up and spin-down particles. Consequently, $\mathcal{F}_{\pm \frac{1}{2}\ell m}$ can also be interpreted as the corresponding spin-up and spin-down responses.

Remarkably, the fermionic response function is \emph{real} and nonvanishing for both Kerr \emph{and} Schwarzschild black holes. 
In the latter case, for spin-$\frac{1}{2}$ fields, it simply reduces to 
\begin{align}
\mathcal{F}_{\pm\frac{1}{2}\ell m}^{\rm Schw}&=\pm 4^{-2 \ell-1}\,,
\end{align}
independent of the azimuthal number $m$, as expected. 

Furthermore, the response for spin-up or spin-down perturbations is the same, modulo a sign, while in the bosonic case $\mathcal{F}_{slm}$ does not depend on the sign of $s$.
Finally, the response is finite for extremal black holes,
\begin{equation}
   \mathcal{F}_{\pm\frac{1}{2}\ell m}^{\,\rm extremal}= \pm 4^\ell m^{2\ell+1}\frac{(\ell+1/2)! (\ell-1/2)!}{(2 \ell)! (2 \ell+1)!}\,, \label{responseExt}
\end{equation}
and grows as $\exp[2\ell (1-\ln 2)]$ in the large $\ell=m$ limit. This exponential growth also persists for near-extremal black holes. Indeed, one can verify that at large $\ell=m$, $ \mathcal{F}_{\pm\frac{1}{2}\ell m}$ behaves as $e^{2\ell x}$ where $x>0$ for $a>a_c \sim 0.95M$, and $x<0$ otherwise.  This implies a sharp distinction in the response to large-$\ell$ perturbations between highly spinning black holes and those with lower spin, a point to which we will return in the conclusion.

\begin{figure}[t!]
	\centering
 	\includegraphics[width=0.48\textwidth]{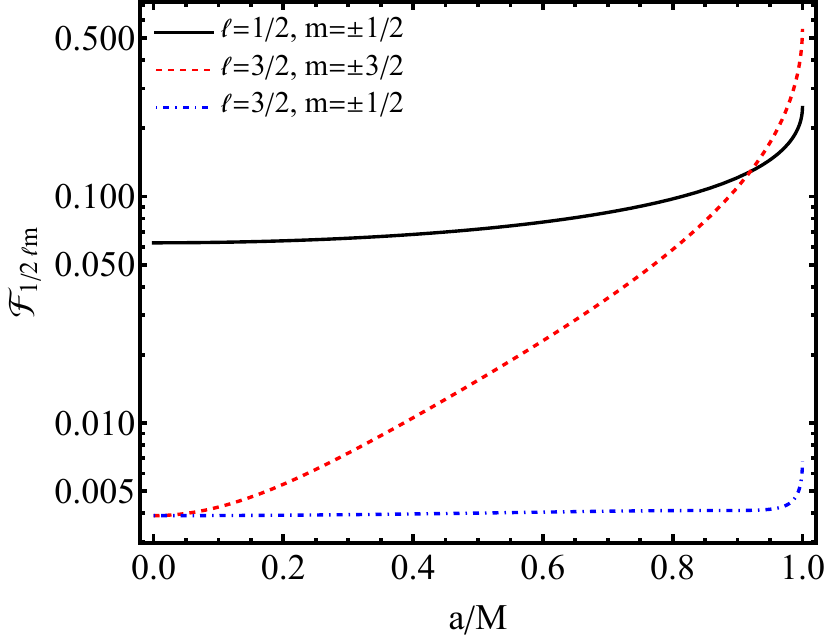}
	\caption{Fermionic Love number of a Kerr black hole to massless spin $s=1/2$ perturbations as a function of the black-hole dimensionless angular momentum for some values of $(\ell,m)$.
    Note that $\mathcal{F}_{\pm\frac{1}{2}\ell m}$ is finite in the extremal ($a\to M$) limit and depends on $m$ only through $m^2$.}
	\label{fig}
\end{figure}

The dependence of the spin-$\frac{1}{2}$ fermionic response on the black-hole angular momentum is shown in~\ref{fig} for smaller values of $(\ell,m)$, which shows the finite extremal value, as well as the $m$-independent Schwarzschild value.

Interestingly, the fact that the response in~\ref{response} is purely real shows that the fermionic dissipation numbers vanish identically for static perturbations. This is another striking difference with respect to the bosonic sector, wherein $\Im({\cal F}_{s\ell m})\propto(\omega-m\Omega_H)$ to linear order in the frequency of the perturbation, where $\Omega_H=a/(2Mr_+)$ is the black hole angular velocity, which provides a dissipative term also in the static ($\omega\to0$) limit. This difference is directly linked to the absence of black hole superradiance for fermions~\cite{Unruh:1973bda,Brito:2015oca}. Indeed, the dissipation numbers are proportional to the energy absorbed at the horizon. For bosonic perturbations the energy flux is proportional to $\omega-m\Omega_H$, hence giving superradiant energy extraction at small frequencies. In contrast, the stress-energy tensor --- and corresponding conserved current --- for Dirac fields imply that the current flowing down the horizon is always positive \cite{Unruh:1973bda}, with the energy flux being proportional to $\omega$, also in the rotating case, thereby vanishing in the static limit.

Even when the fermionic response is extracted from gauge-invariant quantities directly associated with observables, it is well known that within black-hole perturbation theory it remains, subject to a coordinate ambiguity (see, e.g.,~\cite{Pani:2015hfa,Pani:2015nua,Gralla:2017djj}). Indeed, one may perform a coordinate transformation $r\to r(1+A/r^{2\ell+1})$ which shifts the tidal Love number defined in~\ref{psi_4_intermediate} by a quantity proportional to $\mathcal{E}_{s\ell m}A$.
This ambiguity can be resolved by matching the perturbation to an appropriate gauge-invariant coefficient, such as the binding energy of a binary system involving tidally deformed bodies~\cite{Creci:2021rkz}, or the Wilson coefficients of the corresponding effective field theory~\cite{Goldberger:2005cd,Goldberger:2020fot,Porto:2016zng,Hui:2020xxx}.
A simpler solution, adopted here, arises when the perturbations can be computed analytically for arbitrary $\ell$. In this case, one performs an analytical continuation of the angular momentum index $\ell$ and unambiguously identifies the response function from~\ref{psi_4_intermediate}~\cite{Kol:2011vg,LeTiec:2020bos}, since coordinate transformations with $\ell\in\mathbb{C}$ are forbidden. We are therefore confident that our main result does not suffer from any coordinate ambiguity. 

Although the fermionic response is extracted from specific spinor components defined with respect to a chosen tetrad, the Love numbers in~\ref{response} are invariant under spin-frame transformations. Such transformations act as local Lorentz rotations in spin space. As such, they mix the $R_{\pm1/2}$ radial solutions \emph{linearly} but leave unchanged the ratio between the growing and decaying radial modes from which the response coefficients are defined. Hence, the fermionic Love numbers are tetrad- and spin-frame independent. Moreover, parity exchanges the two spin-weight sectors $s\to-s$ (modulo complex conjugation) and leaves the magnitude of the response unchanged, up to the conventional sign associated with spin orientation; so no parity-odd tidal response arises in the static limit, as in the bosonic sector.

Finally, fermionic fields are intrinsically quantum, but \emph{bilinears} computed from them have a classical counterpart. For $s=1/2$, the simplest bilinear one can construct is $\bar\Psi \Psi = \Psi^\dagger \gamma^0 \Psi$. At large distance, using the expressions for $\mathcal{E}_{s\ell m}$ and ${\cal F}_{s\ell m}$, for $s=1/2$ we get
\begin{align}
    \bar\Psi\Psi \underset{r\to \infty}{\sim} r^{2\ell-1} \Big[&(1+{\cal A}+\cdots)\nonumber\\
    &+2{\cal A}{\cal F}_{\frac{1}{2}\ell m}\left(\frac{r_+}{r}\right)^{2\ell+1}(1+\cdots)\Big]\,,
\end{align}
where ${\cal A}^{-1}=\left(\frac{1}{4}+4P^2\right)\left[\left(\ell+\frac{1}{2}\right)^2+4 P^2\right]$. This shows that the nonvanishing fermionic response affects also classical quantities.

{\bf Concluding discussion.}
We have shown that a black hole in General Relativity exhibits a nonzero conservative response to massless fermionic perturbations, providing the first analytic computation of fermionic Love numbers for a generic spin-$s$ perturbation and arbitrary black-hole angular momentum. We also found that the fermionic dissipation numbers vanish for static perturbations, reflecting the absence of superradiance for fermions.  
These findings stand in sharp contrast with the bosonic sector, where all (scalar, electromagnetic, and gravitational) tidal Love numbers of a Kerr black hole vanish, while the dissipation numbers remain proportional to the black-hole angular momentum for static perturbations. 

The vanishing of bosonic tidal Love numbers can be understood in terms of ladder symmetries, whereby generic static $\ell$ modes are related to $\ell=0$ modes through a ladder structure~\cite{Hui:2021vcv,Berens:2022ebl,BenAchour:2022uqo,Katagiri:2023yzm,Sharma:2024hlz}. For $\ell=0$, static solutions contain both growing and decaying components at large distances. However, the ladder symmetry implies that the decaying component --- responsible for the tidal response --- must diverge at the horizon, thereby enforcing vanishing Love numbers for $\ell=0$ and, by symmetry, for all $\ell$. Fermionic perturbations evade this constraint, since their lowest multipole, $\ell=s\in \mathbb{Z}/2$, admits a regular decaying solution. In light of this, it would be interesting to revisit the $SL(2,\mathbb{R})$ ``Love'' symmetry of the near-horizon geometry, which has also been associated with the vanishing of bosonic tidal Love numbers~\cite{Charalambous:2021kcz,Charalambous:2022rre}.

Our results strongly suggest that black holes in General Relativity may generically possess nonvanishing tidal Love numbers under any fermionic perturbations.
This is supported by the fact that spin-$3/2$ test fields in Kerr geometry are governed by the same Teukolsky equations~\ref{teuR}--\ref{teuS} with $s=\pm 3/2$~\cite{Gueven:1980be,TorresdelCastillo:1990aw}; hence, our result in~\ref{response} applies directly to this case.
Since elementary massless fermions with ${\rm spin} >3/2$ are not expected to exist in any consistent relativistic quantum field theory~\cite{Weinberg:1980kq}, our results already encompass all relevant cases. Nonetheless, it may be worthwhile to explore whether \emph{composite} higher-spin excitations can be reformulated in Teukolsky form, in which case our results would apply.

Another natural extension is to charged black holes. In the presence of a background electromagnetic field, the coupling between fermionic and electromagnetic perturbations may produce a new class of nonvanishing mixed fermionic-electromagnetic Love numbers. A related problem has recently been studied for charged scalar perturbations of Kerr–Newman black holes~\cite{Ma:2024few}.

In this letter we focused on the \emph{static} response, but our methods can extend to dynamical settings, generalizing recent analyses~\cite{Bhatt:2024yyz,Katagiri:2022vyz, Ivanov:2022qqt, DeLuca:2023mio,Charalambous:2022rre, Chakraborty:2023zed,Bhatt:2024yyz,Saketh:2023bul,*Perry:2023wmm,*Ivanov:2024sds,*DeLuca:2024ufn,*Katagiri:2024wbg,*Katagiri:2024fpn,*Bhatt:2024rpx,*Caron-Huot:2025tlq,*Chakraborty:2025wvs} to the fermionic sector.

Another promising direction is to embed our computation within the worldline effective field theory~\cite{Goldberger:2005cd, Goldberger:2020fot,Porto:2016zng,Hui:2020xxx}, by analyzing the scattering of massless fermions.

This framework would allow one to quantify the imprint of fermionic Love numbers on the dynamics of black-hole binaries. While this scenario requires one of the companions to source a massless fermionic field, it would open the door to novel phenomenology in gravitational-wave observations, potentially connected to recently discovered black-hole solutions with electroweak hair in Einstein–Weinberg–Salam theory~\cite{Gervalle:2024yxj,*Gervalle:2025awa}.
In particular, it would be interesting to investigate the convergence of the post-Newtonian series in the large $\ell=m$ limit, given that the response of near-extremal black holes grows \emph{exponentially} for both bosonic and fermionic perturbations (see~\ref{responseExt} for Dirac fields in the extremal case but this holds also for the bosonic response in \ref{response}), suggesting that higher-multipole tidal responses could play an unexpectedly significant role when $a\gtrsim 0.95 M$. Such large spins are consistent with estimates for black holes in low-mass X-ray binaries~\cite{Brenneman:2006hw,Middleton:2015osa}, as well as with the spins of the binary components in the recent GW231123 gravitational-wave event~\cite{LIGOScientific:2025rsn}.

Finally, the vanishing of bosonic Love numbers in General Relativity is also tied to classical no-hair theorems, implying that a black hole cannot sustain static bosonic hair decaying at infinity~\cite{Bekenstein:1971hc,*Bekenstein:1972ky,Hui:2021vcv}. Our results highlight the distinctive role of fermions in potentially circumventing these theorems~\cite{Gervalle:2024yxj,*Gervalle:2025awa}, and open new directions to probe the interplay between fundamental fields, black-hole structure, and strong-gravity phenomenology. For example, the nonzero fermionic response indicates that the lowest mode at $\ell=|s|$ corresponds to a static, normalizable mode (see~\ref{psi_4_intermediate}), i.e. a fermionic hair. In particular, for spin-$\tfrac{1}{2}$, we find that the four-spinor does not diverge at infinity while remaining nontrivial at the horizon for $\ell=\tfrac{1}{2}$. The existence of such a mode could have been anticipated from supersymmetric theories of gravity: since General Relativity can be embedded into four-dimensional supergravity, the fermionic superpartners of the bosonic (spin-$1$ and spin-$2$) fields defining the Kerr–Newman solution guarantee the existence of a spin-$\frac{1}{2}$ and spin-$\frac{3}{2}$ normalizable perturbation at the lowest value of $\ell$ in Kerr–Newman. An open question raised by our work is whether nontrivial fermionic hair can also be constructed for the full tower of $\ell$ values for which the response is nonzero, and whether they survive at the nonlinear level.

{\bf Acknowledgments.} 
We thank Miguel Correia, Gerardo Garcia Moreno, and Samir Mathur for interesting comments. 
The research of S.C. is supported by MATRICS (MTR/2023/000049) and Core Research Grants
(CRG/2023/000934) from SERB, ANRF, Government of India.  P.H. is supported by the Department of Physics at The Ohio State University. P.P. is partially supported by the MUR FIS2 Advanced Grant ET-NOW (CUP:~B53C25001080001) and by the INFN TEONGRAV initiative.

\bibliography{References}

\bibliographystyle{./utphys2}

\end{document}